\documentclass[twocolumn]{article}

\usepackage{graphicx}
\usepackage{rotating}
\usepackage{subfigure}

\newcommand{\ie}{\textit{i.e. }}
\newcommand{\eg}{\textit{e.g.}}
\newcommand{\be}{\begin{equation}}
\newcommand{\ee}{\end{equation}}
\newcommand{\nn}{\nonumber \\}
\newcommand{\de}{\stackrel{\mbox{\tiny def}}{=}}

\newcommand{\rmd}{{\mathrm d}}

\newcommand{\real}{\textrm{R}}

\newcommand{\propositionfv}[2]{A_{#1}^{#2}}
\newcommand{\proposition}[2]{#1_{#2}}

\newcommand{\probdist}[3]{\mathrm{p}(#1_{#2}|#3)}
\newcommand{\probdistn}[2]{\mathrm{p}(#1|#2)}
\newcommand{\probdistfv}[3]{p(\propositionfv{#1}{#2}|#3)}
\newcommand{\mipd}[2]{p_m(\propositionfv{#1}{#2})}

\newcommand{\market}[1]{\bar{#1}}

\begin{document}

\title{Managing Derivative Exposure}
\author{U. Kirchner \\ ICAP \\ PO Box 1210, Houghton, 2041, South Africa \\ulrich.kirchner@icap.co.za}

\maketitle

\begin{abstract}
We present an approach to derivative exposure management based on subjective and implied probabilities. We suggest to
maximize the valuation difference subject to risk constraints and propose a class of risk measures
derived from the subjective distribution.

We illustrate this process with specific examples for the two and three dimensional case.
In these cases the optimization can be performed graphically.
\end{abstract}

\section{Introduction}

When using derivatives to create exposure to an underlying asset one important question arises:
Which of the available and how many contracts should be chosen?
Or in other words, which exposure is most desirable given a particular risk profile.

We believe that any proper exposure management has to be based on a subjective probabilistic approach,
incorporating all known information (and uncertainty), not
just historic estimates of statistical parameters.

In the Subjective Approach to Finance \cite{uk-pa} instrument valuations
are based on subjective information and beliefs. These subjective valuations
should then be used together with the current market prices to make
investment decisions expressing how ones views differ from the market implied distributions\cite{uk-mip}.

\section{Notation}
The probability of proposition $A$ to be true, given that $B$ is true is denoted by $P(A|B)$. $A$ and $B$ can be composed of several
propositions. $I$ denotes the available (subjective) background information.

Similarly, probability distributions are denoted by $p(\proposition{A}{x}|B)$, where $\proposition{A}{x}$ is a proposition involving a continuous variable $x$.

Here we will be particularly concerned with the asset price distribution at future times. We use the notation
$\propositionfv{x}{t}$ to represent the statement ``The asset price will be $x$ at time $t$''.
We may omit $t$ if it is clear from the context.

\section{A generic procedure}

In \cite{uk-pa} a generic approach to exposure management in the subjective and probabilistic frame-work was presented,
which is developed here under the assumption
that all parameters except the final asset value are known (in particular the interest rate and dividend yield).
Furthermore, we assume here that valuations and probability distributions are independent of the traded quantity.

Let $\mipd{x}{t}$ be the implied market distribution for the value of an asset at time $t$ --- see \eg \cite{uk-mip} on how
it is defined and how it can be found. This market-implied distribution is determined through, and hence compatible with, current market prices.

Our subjective beliefs about the final value are described by a probability distribution $\probdistfv{x}{t}{I}$, where $I$ indicates
the information available to us. This probability distribution determines our valuations and perceived risks.

Let us consider a set of $N$ (European) derivatives (over the same asset) with pay-off profiles $f_i(x)$ (where $x$
is the final asset value). The current market price is then
\be
\market{V}_i = e^{-rt} \int_0^\infty \mipd{x}{t} \; f_i(x) \; \rmd x
\ee 
and the subjective value of each instrument to us is
\be
V_i = e^{-rt} \int_0^\infty \probdistfv{x}{t}{I} \; f_i(x) \; \rmd x.
\ee

The current portfolio market value for a portfolio containing $n_i$ contracts of instrument $i$ is
then given by
\be
\market{\Pi}(n_1,\ldots,n_N) = \sum_{i=1}^N n_i \market{V}_i.
\label{eq-100}
\ee
On the other hand, our subjective valuation of this portfolio is
\be
\Pi(n_1,\ldots,n_N) = \sum_{i=1}^N n_i V_i.
\label{eq-200}
\ee

We argue that an investor should aim to maximize the valuation difference 
\be
\xi(n_1,\ldots,n_N)=\Pi-\market{\Pi},
\ee
subject to risk constraints.
Note that $\xi$ is a linear function of the $n_i$.

A slightly degenerate case of this is the
``market knows best'' case, where we adopt the market implied distribution as the subjective probability distribution.
Hence the valuation difference vanishes and we optimize risk parameters.

\section{Possible risk measures}
With $x$ being the final asset value, let $\Pi_f(x,n_1,\ldots,n_N)=\sum_{i=1}^N n_i f_i(x)$ be the final portfolio value at expiry. The profit/loss
is then
\be
L(x,n_1,\ldots,n_N) \de \Pi_f - e^{rt} \market{\Pi}
\ee
We are interested in risk measures related to the profit/loss probability distribution $\probdist{L}{l}{I}$,
where $L_l$ is the proposition that the profit/loss is $l$ (in currency terms).

Let us define the loss function
\be
g(l) \de \max(-l,0).
\ee
For each value of $i$ the quantity\footnote{Here we set $g^0(l) \de H(-l)$, where $H$ is the Heaviside step function.}
\be
\rho_i(n_1,\ldots,n_N) \de \int_{-\infty}^\infty \probdist{L}{l}{I} \; g^i(l) \; \rmd l
\ee
can be seen as a risk measure.
The higher $i$ the more the measure penalizes larger losses compared to smaller losses. 

For example, $\rho_0$ is the probability of a realized loss at expiry and $\rho_1$ is the expected value of the loss.

From the linearity of the portfolio values (\ref{eq-100}) and (\ref{eq-200}) it follows that
\be
L(x,\lambda n_1,\ldots,\lambda n_N)
= \lambda L(x,n_1,\ldots,n_N),
\ee
\ie the profit/loss scales linearly with the number of contracts. 
Let $L'_{\lambda l}$ be the proposition that $L(x,\lambda n_1, \ldots, \lambda n_N)$ has the value $\lambda l$. 
For $\lambda \in \real^+$ we have
\be
\probdist{L'}{\lambda l}{I} \rmd (\lambda l) = \probdist{L}{l}{I} \rmd l
\ee
\be
g^j(\lambda l) = \lambda^j g^j(l)
\ee
and hence
\be
\rho_i(\lambda n_1,\ldots,\lambda n_N)
= \lambda^i \rho_i(n_1,\ldots,n_N).
\ee

This corresponds to a simple radial dependence in an N-dimensional spherical coordinate system.
Hence it is advantageous to consider the optimization problem in terms of the exposure radius
\be
n = \sqrt{\sum_{i=1}^N n_i^2}
\ee
and the exposure angles $\alpha_1,\ldots,\alpha_{N-1}$, which
correspond to $N$-dimensional spherical coordinates for $n_1,\ldots,n_N$.

To maximize the valuation difference one proceeds as follows:
For each value of the exposure angles one evaluates all $\rho_j$ for which constraints are imposed.
If $\rho_0$ does not satisfy an imposed constraint, then these exposure angles are not feasible (because $\rho_0$ is independent of the exposure for $n>0$).
For all other constraints evaluate the maximum possible exposure
\be
n_j^\mathrm{max}
=\left(
\frac{\rho_j^\mathrm{max}}{\rho_j(\alpha_1,\ldots,\alpha_{n-1})}
\right)^{1/j}
\ee
and select the smallest.

As the risk measure is based on the subjective distribution function it is important
that it does not understate risks or overstate information. It must
be a rational judgement on the likelihood of all possible outcomes.
Expressing ones information and beliefs in a subjective probability distribution
is in itself a challenging task and will be the subject of further research.

\section{The two-instrument case}

This is the simplest possible case for two derivatives with number of contracts $n_1$ and $n_2$.
Let us introduce polar coordinates by setting
\be
n \de \sqrt{n_1^2 + n_2^2}
\ee
and defining the {\em exposure angle} $\alpha \in [0,\pi)$ such that\footnote{$\alpha$ is ill-defined for $n=0$, but this case is easily treated separately as it corresponds to no exposure at all.}
\be
n_1=n \cos(\alpha) \qquad n_2=n \sin(\alpha).
\ee

Given risk constraints in form of maximum values for $\rho_i$ (for several $i$) we can
calculate a maximum allowed $n$ for each exposure angle $\alpha$.
Furthermore, for each exposure angle $\alpha$ we can calculate the valuation difference $\xi$ for this maximum possible exposure.
These quantities can then be plotted against the exposure angle $\alpha$ to identify the optimal exposure parameters.

\begin{figure}
\begin{center}
\includegraphics[width=3.2in]{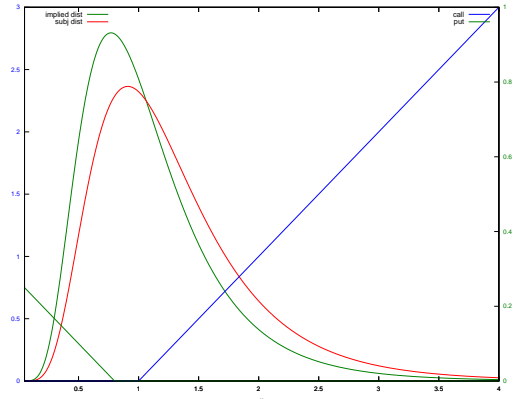}
\end{center}
\caption{
Pay-off profiles of the two instruments and implied (blue) and subjective (red) probability distributions.
}
\label{fig-10}
\end{figure}

An example is presented below in figures \ref{fig-10} and \ref{fig-20}.
For the implied and subjective probability distribution we use two uncertain variance distributions (see Appendix \ref{sec-A1}) 
with a log-normal distribution for the variance.
Both distributions differ in expected return (higher in subjective distribution).
The two instruments considered here are the $100\%$ call and the $80\%$ put.
The first figure shows the implied and subjective probability distribution together with the pay-off
profile of the two instruments. 

The second figure shows the maximum exposure and valuation differences for the constraint $\rho_1 \ge -0.1$. 
From the graph we see that the maximum valuation difference is achieved for $\alpha \approx 286\deg$ with an exposure of $n=2.38$.
This corresponds to the following number of contracts:
\begin{eqnarray}
n_1&=& 2.38 \cos(286 \deg) = 0.66\nn
n_2&=& 2.38 \sin(286 \deg) = -2.29\nonumber
\end{eqnarray}
Hence one should short $2.29$ puts (80\% strike) and buy $0.66$ calls (100\% strike).

\begin{sidewaysfigure*}
\begin{center}
\includegraphics[width=8in]{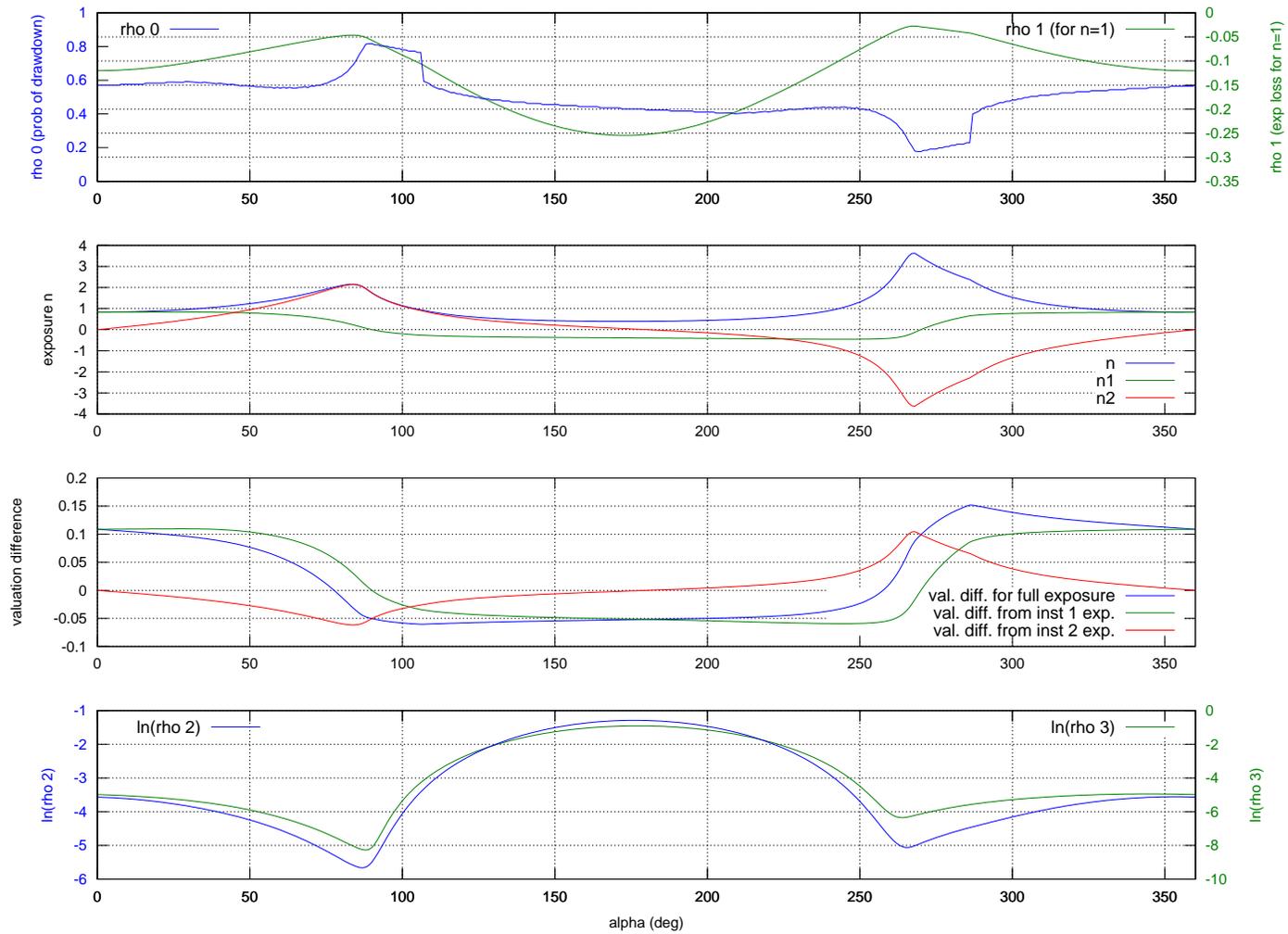}
\end{center}
\caption{
The first graph shows the risk measure $\rho_0$ and $\rho_1$ for $n=1$.
The second graph shows the maximum exposure for $\rho_1(n)\ge -0.1$, which corresponds to
$\rho_1(n)=0.1$.
The third graph then shows the achievable valuation difference under this constraint.
The last graph shows the logarithm of the risk measures $\rho_2$ and $\rho_3$ for $n=1$. Here no constraint has been imposed on these two measures. However, large values indicate
higher tail risks. 
}
\label{fig-20}
\end{sidewaysfigure*}

\section{Three instrument case}

In the three dimensional case we can use ordinary spherical coordinates defined through
\begin{eqnarray}
n_1 &=& n \cos(\alpha_1)
\nn
n_2 &=& n \cos(\alpha_2) \sin(\alpha_1)
\nn
n_3 &=& n \sin(\alpha_2) \sin(\alpha_1) \nonumber
\end{eqnarray}
where $\alpha_1 \in [0,\pi]$ and $\alpha_2 \in [0,2 \pi)$.
This implies
\be
n = \sqrt{n_1^2 + n_2^2 + n_3^2}.
\ee
Note that in this parameterization the two instrument case (for instruments 2 and 3) is recovered for $\alpha_1=\pi/2$.

The maximum achievable valuation difference is now a function of the exposure angles $\alpha_1$ and $\alpha_2$. This function can be plotted as a heat or contour map
for graphical optimization.

\begin{figure}
\begin{center}
\includegraphics[width=3.2in]{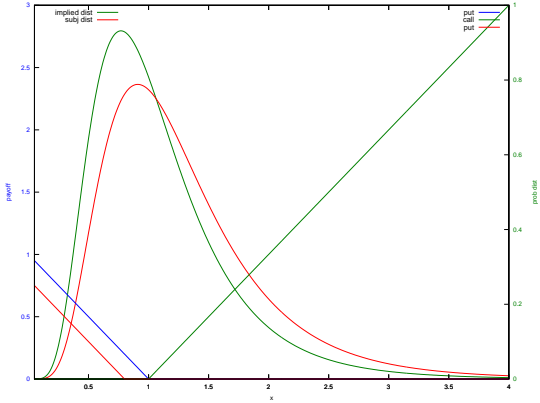}
\end{center}
\caption{
Pay-off profiles of the three instruments and implied (blue) and subjective (red) probability distributions.
}
\label{fig-30}
\end{figure}

\begin{sidewaysfigure*}
\begin{center}
\subfigure[Maximum achievable valuation difference]{
\label{fig-40-a}
\includegraphics[width=3.7in]{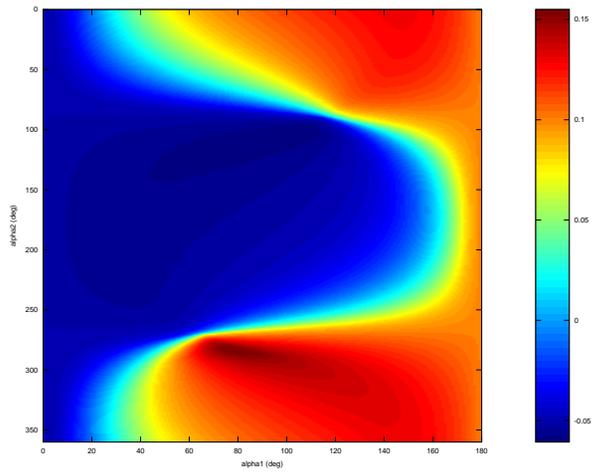}
}
\subfigure[Exposure for fixed $\rho_1$]{
\label{fig-40-b}
\includegraphics[width=3.6in]{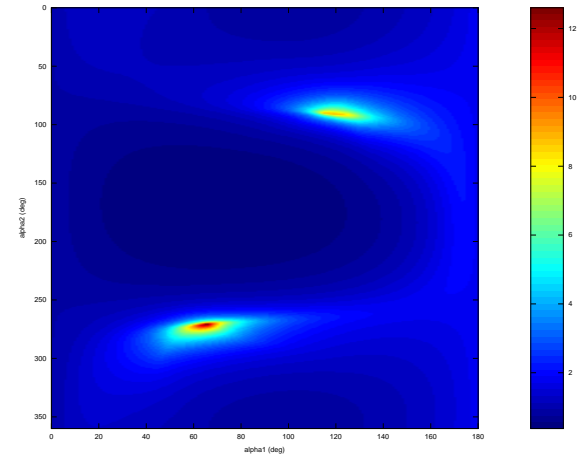}
}
\end{center}
\caption{
Valuation difference and exposure for the three instrument case.
The maximum valuation difference is achieved for $\alpha_1\approx80\deg$, $\alpha_2\approx284\deg$
with an exposure of $n\approx3.14$.
}
\label{fig-40}
\end{sidewaysfigure*}

Figures \ref{fig-30} and \ref{fig-40} extend the previous example to the three instrument case by adding the $100\%$ put (and leaving the distributions unchanged).

From figure \ref{fig-40-a} we can see that the maximum valuation difference is achieved for
\be
\alpha_1 \approx 80 \deg
\qquad
\alpha_2 \approx 284 \deg
\ee
with an exposure of $n\approx3.14$. This corresponds to $n_1=0.55$, $n_2=0.75$, and $n_3=-3$.

Note how adding an extra instrument improves the achievable valuation difference from the two dimensional case.
In fact, as the two dimensional case is a subset of the considered exposures this must be the case. 

\section{Generic $N$-dimensional case}
 While it is more difficult to represent this case graphically, one can still follow the generic procedure.
We introduce hyper-spherical coordinates
\begin{eqnarray}
n_1 &=& n \cos(\alpha_1)
\nn
n_j &=& n \cos(\alpha_j) \Pi_{k=1}^{j-1} \sin(\alpha_k)
\nn
n_N &=& n \Pi_{k=1}^{N-1} \sin(\alpha_k) \nonumber
\end{eqnarray}
where
\begin{eqnarray}
\alpha_j &\in& [0,\pi] \mbox{ for } 1\le j < N \nn
\alpha_N &\in& [0,2\pi) \nonumber
\end{eqnarray}
As there are now at least three exposure angles involved, the optimization cannot simply be performed graphically in the parameter space.
However, there are a number of options.
Firstly, brute force computation of all possible options (on a discrete grid) and selection of the optimum may be viable for lower
resolution and dimensionality.
Secondly, a genetic algorithm could be used.

\section{Conclusion}

We presented a generic approach to derivative exposure management for one underlying asset.
We suggest to maximize the valuation difference (between subjective and market valuation) under risk constraints.

A class of risk measures based on the expected value of powers of the loss function was introduced.
Higher powers of the loss function penalize more for the possibility of higher losses. 

We illustrated this procedure for the two and three instrument case.
In these cases the optimization can be performed graphically.
One plots the maximum achievable valuation difference
for each combination of exposure angles and selects the highest value
compatible with any other potential risk constraints (which can be plotted in similar graphs).

Above approach can in principle be generalized to multiple underlying assets. However, the interdependence cannot be ignored
and hence a subjective multivariate distribution is needed\footnote{The implied distribution only
enters through the known current asset prices, and hence takes implied interdependencies automatically into account.}. 

The presented approach relies on the availability of a subjective probability distribution.
However, to translate information and beliefs into such a distribution is in itself a non-trivial exercise and
will be the subject of further research.

\appendix
\section{Uncertain variance distributions}
\label{sec-A1}
Given maximum-entropy arguments it is reasonable to assume a Gaussian (normal) distribution if the mean and variance $\sigma^2$ are known.
Let us denote this normal distribution for the log-returns $l$ by
\[
\probdistn{\proposition{L}{l}}{\proposition{v}{\sigma^2} \proposition{M}{\mu} I}= \frac{1}{\sqrt{2 \pi} \sigma} \exp\left(-\frac{(l-\nu(\mu,\sigma))^2}{2 \sigma^2}\right),
\]
where $\proposition L l$ is the proposition that the log return will be $l$, $\proposition{v}{\sigma^2}$ is the proposition that the variance has the value $\sigma^2$,
$\proposition{M}{\mu}$ is the proposition that the mean of the returns (not log-returns!) is $\mu-1$, and
the log-mean $\nu$ is
\be
\nu(\mu,\sigma) \de \ln(\mu)-\sigma^2/2.
\ee  

If the variance is not known we have after marginalizing
\[
\probdistn{\proposition{L}{l}}{\proposition{M}{\mu} I}
= \int_0^\infty \probdistn{\proposition{L}{l}}{\proposition{v}{\sigma^2} \proposition{M}{\mu} I}
 \probdistn{\proposition{v}{\sigma^2}}{\proposition{M}{\mu} I} \; \rmd \sigma^2, \nonumber
\]
where the second factor describes our beliefs about the value of the variance.

The mean log return (the mean return is by construction $\mu-1$) is then given by
\begin{eqnarray}
\lambda &\de& \int_{-\infty}^\infty \rmd l\;  l \; \probdistn{ \proposition L l}{\proposition M \mu I} \nn
&=& \int_0^\infty \rmd \sigma^2 \int_{-\infty}^\infty \rmd l \; l \;\probdistn{\proposition L l}{\proposition v {\sigma^2} \proposition M \mu I} \;
\probdistn{ \proposition v \sigma^2}{I} \nn
&=& \int_0^\infty \rmd \sigma^2 \probdistn{ \proposition v \sigma^2}{I}  \; \nu(\mu,\sigma) \nn
&=& \ln(\mu) - \bar{\sigma^2}/2,
\end{eqnarray}
where we defined the average variance as
\be
\bar{\sigma^2} \de  \int_0^\infty \rmd \sigma^2 \; \sigma^2 \; \probdistn{ \proposition v \sigma^2}{I}
\ee

The variance is the second central moment of the log return distribution. Hence
\begin{eqnarray}
\xi &=& \int_{-\infty}^\infty \rmd l (l-\lambda)^2 \probdistn{\proposition L l}{\proposition M \mu I}
\nn
&=& \int_{-\infty}^\infty \rmd l  (l^2 - 2\lambda l + \lambda^2)\probdistn{\proposition L l}{\proposition M \mu I}
\nn
&=& - \lambda^2 +  \nn
&&\int_0^\infty \rmd \sigma^2 \;
\probdistn{\proposition v \sigma^2}{I}
\underbrace{\int_{-\infty}^\infty \rmd l \; l^2 \;\probdistn{\proposition L l}{\proposition v {\sigma^2} \proposition M \mu I}}_{=\sigma^2+\nu^2}
\nn
&=& \bar{\sigma^2} +(\bar{\sigma^4}-\bar{\sigma^2}^2)/4,
\end{eqnarray}
where
\be
\bar{\sigma^4} \de \int_0^\infty \rmd \sigma^2 \; \sigma^4 \;
\probdist{v}{\sigma^2}{I}.
\ee

Let us consider the case where $\probdist{v}{\sigma^2}{I}$ is a log-normal
distribution for $\sigma^2$ with uncertainty parameter $\beta>0$
\[
\probdist{v}{\sigma^2}{I}
=\frac{1}{\sigma^2 \beta \sqrt{2 \pi}}
\exp\left(
- \frac{(\ln(\sigma^2)-\ln(\alpha))^2}{2 \beta^2}
\right)
\]
Then with $\gamma\de e^{\beta^2/2}$
\be
\bar{\sigma^2}
=\alpha \gamma
\ee
\be
\bar{\sigma^4} = \exp\left(
2 \ln(\alpha) + \frac 1 2 4 \beta^2
\right)
=\alpha^2 \gamma^4.
\ee
Hence the mean log return becomes
\be
\lambda^* = \ln(\mu) - \frac{\alpha \gamma}{2},
\ee
and the variance
\be
\xi^* =\alpha \gamma +\frac{\alpha^2 \gamma^2}{4}(\gamma^2 -1).
\ee

\end{document}